\documentclass[pra,aps,showpacs,superscriptaddress]{revtex4}

\usepackage{amsmath}
\usepackage{epsfig}
\usepackage{graphicx}
\usepackage{dcolumn}
\usepackage{bm}

\begin{document}


\title{Stability and chaotic behaviors of Bose-Einstein condensates in optical lattices with two- and three-body interactions}

\author{Yan Chen}
\affiliation{Institute of Theoretical Physics, Lanzhou University, Lanzhou $730000$, China}

\author{Ke-Zhi Zhang}
\affiliation{Physics and Electronics Engineering College, Northwest Normal University, Lanzhou 730070, China}

\author{Yong Chen}
\altaffiliation{Corresponding author. Email: \texttt{ychen@gmail.com}}
\affiliation{Institute of Theoretical Physics, Lanzhou University, Lanzhou $730000$, China}
\affiliation{Key Laboratory for Magnetism and Magnetic Materials of the Ministry of Education, Lanzhou University, Lanzhou 730000, China}

\date{\today}

\begin{abstract}
The stability and chaotic behaviors of Bose-Einstein
condensates with two- and three-atom interactions in optical
lattices are discussed with analytical and numerical methods. It
is found that the steady-state relative population  appears
tuning-fork bifurcation when the system parameters are changed to
certain critical values. In particular, the existence of
three-body interaction not only transforms the bifurcation point
of the system but also affects greatly on the  macroscopic quantum
self-trapping behaviors of the system associated with the
critically stable steady-state solution. In addition, we also
investigated the influence of the initial conditions, three-body
interaction and the energy bias on the macroscopic quantum
self-trapping. Finally, by applying the periodic modulation on the
energy bias, we find that the relative population oscillation
exhibits a process from order to chaos, via a series of
period-doubling bifurcations.
\end{abstract}

\maketitle

\section{Introduction}
In Recent years, Bose-Einstein condensates (BECs) in optical
lattices have attracted enormous attention both experimentally and
theoretically~\cite{ce1,ce2}. This is mainly because the lattice
parameters and interaction strength can be manipulated using a
modern experimental technique. Making use of this, researchers
have discovered many long-predicted phenomena, for example
non-linear Landau-Zener tunneling, energetic and dynamical
instability and the strongly inhibited transport of
one-dimensional BEC in optical
lattices~\cite{ce3,ce4,ce5,ce6,ce7,ce8,ce9,ce10}. More attracting
phenomena, namely, self-trapping, was recently observed
experimentally in this system~\cite{ce11}. In such an experiment,
a BEC cloud with repulsive interaction initially loaded in optical
lattices was self-trapped. Many theoretical analysis was also
presented about self-trapping~\cite{ce12,ce13,ce14,ce15}. It is
well know the macroscopic quantum self-trapping (MQST) means
self-maintained population imbalance with non-zero average value
of the fractional population imbalance which was detailed
discussed~\cite{ce16,ce17}. Marino et. al. considered that the
damping decays all different oscillations to the zero-phase
mode~\cite{ce18}. Besides, macroscopic quantum fluctuations have
also been discussed by taking advantage of second-quantization
approaches~\cite{ce19}. However, when the trapping potential is
time dependent and the damping and finite-temperature effect can
not be neglected, chaos emerges. Abdullaev and Kraenkel studied
the nonlinear resonances and chaotic oscillation of the fractional
imbalance between two coupled BEC's in a double-well trap with a
time-dependent tunneling amplitude for different
damping~\cite{ce20}. When the asymmetry of the trap potential is
time-dependent and  its amplitude is so small that can be took as
a perturbation, Lee et al. studied the chaotic and
frequency-locked atomic population oscillation between two coupled
BECs with a weak damping, and discovered that the system comes to
an stationary frequency-locked atomic population oscillations from
transient chaos~\cite{ce21}.

It is important to note that theoretical studies of stability are
mainly focused on the effect of two-body interactions. It is clear
that in low temperature and density, where interatomic distance is
much greater than the distance scale of atom-atom interactions,
two-body s-wave scattering should be important and three-body
interactions can be neglected. But, if the atom density is higher,
for example, in the case of BEC in optical lattices, three-body
interactions will play an important role~\cite{ce22}. As reported
in Ref.~\cite{ce23},  even for a small strength of the three-body
force, the region of stability for the condensate can be extended
considerably.

Therefore, the purpose of this paper is to investigate the
steady-state solution of BEC in an one-dimensional periodic
optical lattice when both the two-body and three-body interactions
are taken into account. By using the mean-field approximation and
linear stability theorem, one interesting result is found that the
tuning-fork bifurcation of steady-state relative population
appears when the system parameters are changed to certain critical
values.  The existence of three-body interaction not only
transforms the bifurcation point of the system but also affects
greatly on the self-trapping behaviors of the system associated
with the critically stable steady-state solutions. Additionally,
we also study the effects of the initial conditions, three-body
interaction and the energy bias on the MQST. Besides, we discuss
the chaos behaviors of the system by applying the periodic
modulation on the energy bias. The result shows the relative
population oscillation can undergo a process from order to chaos,
via a series of period-doubling bifurcations.

This paper is organized as follows. In Sec. II, we introduce the
mean-field description of BEC in optical lattices with two- and
three-atom interactions. In Sec. III, with linear stability
theorem, we analysis the stability of steady-state solutions. Then
the influences of three-body interaction on  the macroscopic
quantum self-trapping of the system are displayed In Sec. IV. In
Sec. V, by applying the periodic modulation in the energy bias, we
discuss chaotic behaviors of the system using the numerical
simulation method. In the last section, summary and conclusion of
our work are presented.

\section{Mean-field description of BEC in optical lattices with two- and three-atom interactions}

We focus our attention on a BEC with both two- and three-body
interactions is subjected to one dimensional
 (1D) optical lattices where the motion in the perpendicular
directions is confined. In the mean-field approximation , the
dynamics of BEC can be modeled by the 1D Gross-Pitaevskii (GP)
equation in the comoving frame of the
lattice~\cite{ce3,ce6,ce24,ce25},
\begin{equation}\label{01}
i\hbar\frac{\partial\Phi}{\partial t} = -\frac{1}{2m} \left( \hbar\frac{\partial}{\partial
t}-ima_{l}t \right) ^2 \Phi+\upsilon_0\cos(2K_{l}x)\Phi +\frac{2\hbar^{2}a_s}{a_{\bot}^{2}m}|\Phi|^2\Phi+ \frac{g_2}{3\pi^2a_\bot^4} |\Phi|^4 \Phi,
\end{equation}
where  $\Phi$ is the wave function of the condensate, $m$ is the
mass of atoms, $a_s$ is the two-body s-wave scattering length,
$\upsilon_0$ is the strength of the periodic potential, $K_l$ is
the wave number of the laser light which is used to generate the
optical lattice, $ma_l$ stands for either the inertial force in
the comoving frame of an accelerating lattice or the gravity
force, $a_\bot=\sqrt{\hbar/(m\omega_\bot)}$, where $\omega_\bot$
is the radial frequencies of the anisotropic harmonic trap,
$g_2|\Phi|^4\Phi/(3\pi^2a_\bot^4)$ is three-body interactions
related to the GP equation. Among Eq.~(\ref{01}), all the
variables can be rescaled to be dimensionless by the following
system's basic parameter
$x\sim2K_lx,\Phi\sim\frac{\Phi}{\sqrt{2K_lN}},t\sim\frac{4\hbar}{m}K_l^2t$.
we obtain the normalized 1D-GP equation in optical lattices with
cubic and quintic nonlinearities,
\begin{equation}\label{02}
i\frac{\partial\Phi}{\partial t}=-{{1}\over
{2}} \left( \frac{\partial}{\partial t}-i\alpha t \right)^2\Phi + \upsilon \cos(x)\Phi + c|\Phi|^2\Phi + \lambda|\Phi|^4\Phi,
\end{equation}
where  $\upsilon=\frac{m\upsilon_0}{4\hbar^2K_l^2},
\alpha=\frac{m^2}{8\hbar^2K_l^3}a_l,  c=\frac{Na_s}{K_la\bot^2}
\nonumber$ is the effective two-body interaction, $N$ is the total
numbers of atoms,
$\lambda=\frac{mg_2N^2}{3\pi^2\hbar^2a_\bot^4}\nonumber$ is the
effective interaction among three atoms, here the three-body
interaction is expected to be positive with a value of
$0<\lambda<1$.

In the neighborhood of the Brillouin Zone edge $k=1/2$,
the wave function can be approximated by~\cite{ce3}
\begin{equation}\label{03}
\Phi(x,t)=a(t)e^{ikx}+b(t)e^{i(k-1)x},
\end{equation}
where $a(t)$, $b(t)$ are the probability amplitudes of atoms in
each of the two wells respectively and $|a|^2+|b|^2=1$. By
inserting such wave functions into Eq.~(\ref{02}) and performing
some spatial integrals, we obtain the dynamical equations with
two- and three-body interactions.
\begin{eqnarray}
i\frac{\partial a}{\partial t} &=& \frac{\gamma}{2}a + \frac{c}{2} \left( |b|^2-|a|^2 \right) a+\lambda \left( 1+2|a|^2|b|^2+2|b|^2 \right) a+\frac{\upsilon}{2}b, \label{04}\\
i\frac{\partial b}{\partial t} &=& -\frac{\gamma}{2}b - \frac{c}{2} \left( |b|^2-|a|^2 \right)b + \lambda \left( 1+2|a|^2|b|^2+2|b|^2 \right)b+\frac{\upsilon}{2}a. \label{05}
\end{eqnarray}
Here, the level bias $\gamma(t)=\alpha t$, and $\alpha$ is the
sweeping rate, $c$ and $\lambda$ represent the nonlinear
parameters, $\upsilon$ is the coupling constant between the two
condensates. We introduce the relative population variance
\begin{equation}\label{06}
s=|b|^2-|a|^2,
\end{equation}
with the parameters $a=|a|e^i\theta_a$, $b=|b|e^i\theta_b$,
\begin{equation}\label{07}
\theta=\theta_b-\theta_a.
\end{equation}
Combining Eqs.~(\ref{04}-\ref{07}), one yields the equations of
the relative population and relative phase,
\begin{eqnarray}
\dot{s}&=&-\upsilon\sqrt{1-s^2}\sin\theta, \label{08} \\
\dot{\theta}&=&\gamma+(c+2\lambda)s+\frac{\upsilon
s}{\sqrt{1-s^2}}\cos\theta. \label{09}
\end{eqnarray}
$\dot{s}$ and $\dot{\theta}$ denote the time derivative of the
relative population and the relative phase. If we regard $s$ and
$\theta$ as the canonically conjugate variables Eqs. (8) and (9),
become a pair of Hamilton's canonical equations with the conserved
effective Hamiltonian
\begin{equation}\label{10}
H=\gamma s + \frac{1}{2} (c+2\lambda)s^2 + \upsilon\sqrt{1-s^2} \cos\theta.
\end{equation}
In the following section, we will discuss the stability of
steady-state in the symmetric condition ($\gamma=0$) with linear
stability theorem.

\section{Stability analysis of the steady-state solutions}

In Sec. II, we have given the dynamical equations of the system
with three-body interaction. In this section, we will discuss the
stability of steady-state in the symmetric condition. Generally,
there are two ways to study the stability of nonlinear system,
 the linear stability theorem and the Lyapunov direct
method. We will investigate the stability of the system with the
first method.

 The steady-state solution of this system can be obtained by
setting Eqs.~(8) and (9) to zero. The forms of steady-state
solutions are very complicated when the level bias $\gamma \neq0$.
For simplicity, we set $\gamma =0$, leading to
 \begin{eqnarray}
 \dot{s}&=&f_1(s,\theta)=-\upsilon\sqrt{1-s^2}\sin\theta, \label{11} \\
 \dot{\theta}&=&f_2(s,\theta)=(c+2\lambda)s+\frac{\upsilon
 s}{\sqrt{1-s^2}}\cos\theta. \label{12} \\
 \end{eqnarray}
 and the conserved energy
\begin{equation}\label{13}
 H=\frac{1}{2}(c+2\lambda)s^2+\upsilon\sqrt{1-s^2}\cos\theta.
 \end{equation}
 Taking $\dot{s}=0$, $\dot{\theta}=0$, we get
 \begin{eqnarray}
 -\upsilon\sqrt{1-s^2}\sin\theta&=&0, \label{14}\\
(c+2\lambda)s+\frac{\upsilon s}{\sqrt{1-s^2}}\cos\theta&=&0. \label{15}
\end{eqnarray}
The steady-state solutions obeyed Eqs.~(14) and (15) regard as
\begin{eqnarray}
\theta_1 &=& 2n\pi, \quad s_1=0 \quad  \mathrm{for} \quad H=-\upsilon, \label{16} \\
\theta_2 &=& (2n+1)\pi, \quad s_2=0 \quad   \mathrm{for} \quad  H=\upsilon, \label{17}
\end{eqnarray}
\begin{equation}\label{18}
\theta_{3,4}=(2n+1)\pi,\quad
   s_{3,4}=\pm\sqrt{1-(\frac{\upsilon}{c+2\lambda})^2}\quad
  \mathrm{for} \quad  H=\frac{(c+2\lambda)^2+\upsilon^2}{2(c+2\lambda)^2}.
\end{equation}
According to the linear stability theorem, we look for the
perturbed solutions which are near the steady-state solutions,
\begin{equation}\label{19}
s(t)=s_i(t)+\varepsilon_1(t),\qquad
\theta(t)=\theta_i(t)+\varepsilon_2(t)
\end{equation}
where $s_i(t)$, $\theta_i(t)$ for $i=1,2,3,4$ signify the
steady-state solutions, $|\varepsilon_1(t)|\ll|s_i(t)|$ and
$|\varepsilon_2(t)|\ll|\theta_i(t)|$ which is relate to the
first-order perturbed. Inserting the above expression into Eqs.~(11) and (12), we can obtain the linear equations near to the
steady-states of the nonlinear equations as
\begin{equation}\label{20}
\dot{\varepsilon_1}= \left( \frac{\partial f_1}{\partial
s} \right) _1\varepsilon_1+ \left( \frac{\partial f_1}{\partial
\theta} \right)_1\varepsilon_2\qquad
namely\qquad\dot{\varepsilon_1}=a_{11}\varepsilon_1+a_{12}\varepsilon_2
\end{equation}
\begin{equation}\label{21}
\dot{\varepsilon_2}= \left( \frac{\partial f_2}{\partial
s} \right)_2\varepsilon_1+ \left( \frac{\partial f_2}{\partial
\theta} \right)_2\varepsilon_2\qquad
namely\qquad\dot{\varepsilon_2}=a_{21}\varepsilon_1+a_{22}\varepsilon_2
\end{equation}
Now, we make use of the above expression to investigate the
stability of the steady-states of Eqs.~(16-18).

(1)For  $\theta_1=2n\pi, s_1=0,  H=-\upsilon$, we can calculate the
matrix elements
  $a_{11}=0$, $a_{12}=-\upsilon$, $a_{21}=(c+2\lambda)+\upsilon$, $a_{22}=0$. So,
  the coefficient matrix of the linearized equations (20) and (21) becomes
   $A_1=\left[%
\begin{array}{cc}
  {0}&{-\upsilon} \\
  {c+2\lambda+\upsilon}&{0}\\
\end{array}%
\right]$
such that the characteristic equation
writes $\det(A_1-\lambda I)=\left[%
\begin{array}{cc}
  {0-\lambda}&{-\upsilon} \\
  {c+2\lambda+\upsilon}&{0-\lambda}\\\end{array}%
\right]=0$, which reveals
  that $\lambda^2+\upsilon(c+2\lambda+\upsilon)=0$. We solve the
  equation to get the two eigenvalues of the matrix A
  as $\lambda_1=\sqrt{-\upsilon(c+2\lambda+\upsilon)},
  \lambda_2=-\sqrt{-\upsilon(c+2\lambda+\upsilon)}$. In response to the forms of the eigenvalues, there exist two
  cases for the stabilities:

  (a) $\upsilon(c+2\lambda+\upsilon)\geq0$, that is
  \begin{equation}\label{22}
  \upsilon>0\quad and \quad(c+2\lambda)\geq-\upsilon
  \end{equation}
   \begin{equation}\label{23}
   \upsilon<0\quad and \quad(c+2\lambda)\leq-\upsilon
  \end{equation}
so the two eigenvalues are both pure imaginary numbers. Thus, the
stability of the steady-state solutions $(\theta_1,s_1)$
corresponds to a critical case~\cite{ce26} and the dynamical
bifurcations between the unstable and stable steady-states will
appear when the parameters with two- and three-body interactions
are changed.

  (b) $\upsilon(c+2\lambda+\upsilon)<0$, namely
   \begin{equation}\label{24}
  \upsilon>0\quad and \quad(c+2\lambda)<-\upsilon
  \end{equation}
   \begin{equation}\label{25}
   \upsilon<0\quad and \quad(c+2\lambda)>-\upsilon
  \end{equation}
  so the two eigenvalues are real number. It means that
  $\varepsilon_1$ and $\varepsilon_2$  tend to infinity with the
  increase of time, and the steady-state solutions $(\theta_1,s_1)$
  are unstable.

  (2)For $\theta_2=(2n+1)\pi$, $s_2=0$, $H=\upsilon$, the matrix
  elements write as
  $a_{11}=0,a_{12}=-\upsilon,a_{21}=(c+2\lambda)-\upsilon,a_{22}=0$.
  The corresponding eigenvalues of the matrix $A_2$ become
  $\lambda_1=\sqrt{-\upsilon(\upsilon-(c+2\lambda))},
  \lambda_2=-\sqrt{-\upsilon(\upsilon-(c+2\lambda))}$.
Similarly, there are  two cases of the stabilities:

(a) $\upsilon(\upsilon-(c+2\lambda))>0$, that is
\begin{eqnarray}
(c+2\lambda)>0\quad &\mathrm{and} \quad \upsilon>(c+2\lambda) \label{26}\\
(c+2\lambda)<0 \quad &\mathrm{and} \quad \upsilon>0. \label{27}
\end{eqnarray}
so the two eigenvalues are both pure imaginary numbers. And the
stability of the steady-state solutions $(\theta_2,s_2)$ of the
nonlinear equations are reviewed as critical and the dynamical
bifurcations will occur.

(b) $\upsilon(\upsilon-(c+2\lambda))\leq0$, that is
\begin{eqnarray}
\upsilon>0\quad &\mathrm{and}& \quad (c+2\lambda)\geq\upsilon \label{28} \\
(c+2\lambda)<\upsilon \quad &\mathrm{and}& \quad \upsilon<0 \label{29}
\end{eqnarray}
so the two eigenvalues are positive or negative real number,
respectively. $\varepsilon_1$, $\varepsilon_2$ tend to infinity
as increasing the time to infinity, and the steady-state solutions
$(\theta_2,s_2)$ are losing  their stability.

(3)For $\theta_{3,4}=(2n+1)\pi,
s_{3,4}=\pm\sqrt{1-(\frac{\upsilon}{c+2\lambda})^2},
H=\frac{(c+2\lambda)^2+\upsilon^2}{2(c+2\lambda)^2}$, the matrix
elements read
$a_{11}=0$, $a_{12}=\upsilon^2 /(c+2\lambda)$, $a_{21}=(c+2\lambda)-(c+2\lambda)^3/\upsilon^2$, $a_{22}=0$,
and the eigenvalues $\lambda_1=\sqrt{\upsilon^2-(c+2\lambda)^2)}$, $\lambda_2=-\sqrt{\upsilon^2-(c+2\lambda)^2)}$. In Eq.~(18) the
population $s_{3,4}$ are both real quantities which implies
\begin{equation}\label{30}
(c+2\lambda)^2>\upsilon^2
\end{equation}

\begin{figure}
\begin{center}
\includegraphics[width=0.5\textwidth]{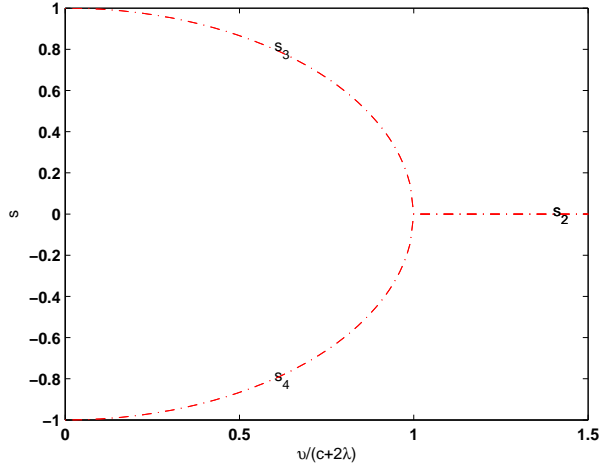}
 \caption{Plots of the
tuning-fork bifurcation from Eqs.~(17) and (18), where
$s_2,s_3,s_4$ are the steady-state solutions and the bifurcation
point is $\frac{\upsilon}{c+2\lambda}=1$} \label{P2}
\end{center}
\end{figure}

Therefore, the two eigenvalues are  pure imaginary numbers. The
stability of the steady-state solutions$(\theta_{3,4},s_{3,4})$ of
the nonlinear equations are regarded as critical and the dynamical
bifurcations will emerge at the bifurcation point
$(c+2\lambda)=\upsilon$, $s=0$. Obviously, the existence of
three-body interaction can change the bifurcation point of the
system. It plays a important role for stability analysis of the
system, as shown in Fig.~\ref{P2}. For
$\frac{\upsilon}{c+2\lambda}>1$, the system is in the critically
stable steady-state ($\theta_2,s_2$), and for
$\frac{\upsilon}{c+2\lambda}<1$, ($\theta_2,s_2$) is unstable and
the two steady-state solutions ($\theta_{3,4},s_{3,4}$) are
critically stable. This is a typical tuning-fork bifurcation, and
the bifurcation point is $\frac{\upsilon}{c+2\lambda}=1$

  According to the above analysis, we conclude that three
steady-state solutions possess different stability for different
parameter regions. And it is very interesting to arrive at the
critically stable steady-state solution in experiment which
relate to the stable stationary MQST~\cite{ce26}. In the following
section, we will illustrate the MQST of the non-stationary states
in detail by two different methods.

\section{The macroscopic quantum
self-trapping of BEC with two- and three-atom interactions}

In this section, we investigate the macroscopic quantum
self-trapping by plotting the phase trajectories and the time
evolution of the relative population of the system.

\subsection{The phase trajectories diagram }

The macroscopic quantum self-trapping refers to the phase space
trajectories whose the relative population is not equal zero. This
can be well understood from the analysis Eqs.~(8)-(10),
corresponding to the critically stable steady-state solutions
discussed in sec.II. Three kinds of cases  occur with different
three-body interaction parameters, as shown in Fig.2.

(1) In the case of $\upsilon=0.2,c=0.1,0<\lambda<0.05$ in the
phase space , there are two stable points $P_1,P_2$ at
$s=0,\theta=\pi$ and $s=0,\theta=0$ respectively [Fig.~2(a)], from
the circumstance described by Eqs.~(22) and (26). Obviously, for
the stable points $P_1$, $P_2$, the atoms distributions are equal in
the two adjacent wells, the relative population of the
trajectories around them is equal to $0$. It means that atoms
oscillate between  two adjacent wells and the macroscopic quantum
self-trapping phenomenon does not emerge in this case.

(2) When parameter is set to
$\upsilon=0.2$, $c=0.1$, $0.05\leq\lambda<0.15$, two more fixed points
emerge in the line $\theta=\pi$ marked by $P_3$, $P_4 $. Among them,
$P_1$, $P_3 $ are steady which is corresponding to condition of
Eq.~(30). They are located in
$s=\pm\sqrt{1-(\frac{\upsilon}{c+2\lambda})^2}$, hence, $P_4$ is
unstable point which lies in $s=0$ and corresponds to condition of
Eq.(26). As seen from Fig.~2(b), for the stable points $P_1,P_3$,
 the atoms distributions are not equilibrium  between two adjacent wells,
and the relative population of the trajectories around them is not
equal to $0$. It indicates that atoms are  self-trapped in one
well. We take it as oscillating-phase-type because the relative
population $s$ and the relative phase $\theta$ oscillate around
the fixed points.

\begin{figure}
\includegraphics[width=1\textwidth]{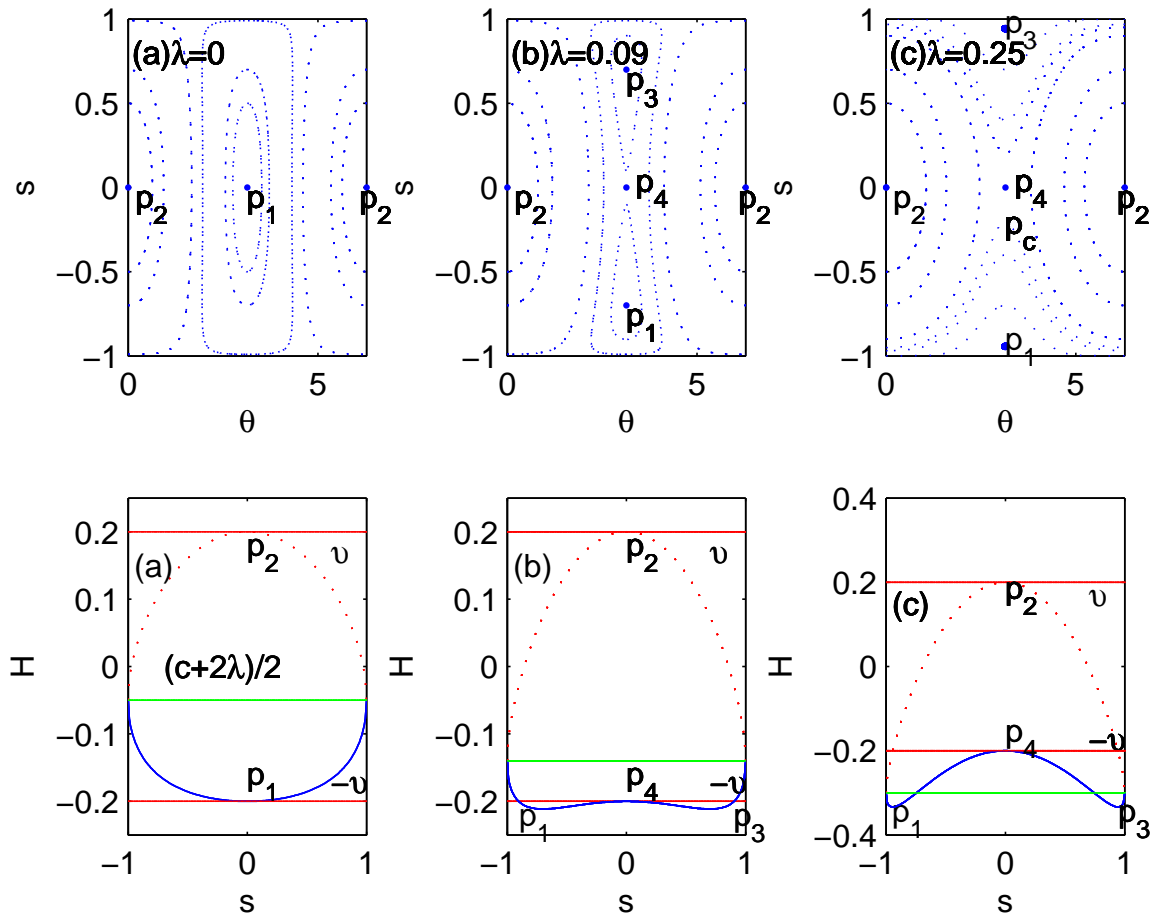}
\caption{Trajectories on the phase space of the system with
three-body interaction varying from $\lambda=0$ to
$\lambda=0.25$(the first row). Corresponding to in the second row
we plot the energy profiles for the relative phase $\theta=0$ (red
dashed) and $\theta=\pi$ (blue solid)}
\end{figure}

(3) For $\upsilon=0.2,c=0.1$, $\lambda\geq0.15$ , It emerges new
trajectories , i.e.the trajectories across point $P_c$
[Fig.~2(c)]. Only the fixed point $P_2$ is stable which is relate
to Eq.~(22). So for these trajectories, $s$ varies with time from
region of $[-1,0]$ to $[0,1]$, Apparently $\langle s\rangle\neq0$,
atoms are self-trapped in one well. We regard it as
running-phase-type macroscopic quantum self-trapping, as described
in Refs.~\cite{ce27,ce28} and observed in experiment~\cite{ce29}.

  The above changes on the topological structure of the phase
space are concerned with the change of the energy profile. When
the relative phase is zero or $\pi$, energy relying on the
parameter with three-body interaction  and the average population
$s$ can be derived from Eq.~(10). Seeing Fig.~2 , the transition
from case(1)to case(2) corresponds to the bifurcation of the
energy profile of $\theta=\pi$: energy curve bifurcates from a
single minimum to the curve of two minima. It means the system
goes from the Rabi regime into the self-trapping regime through
this bifurcation. The lowest order of energy profile with
$\theta=0$ is $-\frac{c+2\lambda}{2}$, and the energy of the
unstable point $P_4$ is $-\upsilon$  which is located on the
maximal order of energy profile with $\theta=\pi$. The results
displayed by the phase space trajectories conform to the case of
steady-state solutions discussed in Sec.III. The transition from
case (2) to case(3) is signified by the overlap of the two energy
regions of the profile. In this condition the trajectory stared
from $s=-1$, $\theta=0$ should be confined to the lower half of
phase plane, corresponding to the running-phase-type macroscopic
quantum self-trapping.

Connecting the analysis of the steady-state solutions to the above
analysis on the energy profile, it concludes that stable
behaviors of the system change constantly with the increase of
$\lambda$ and we obtain a general criterion for the macroscopic
quantum self-trapping trajectories, namely,
$H(s,\theta)<-\upsilon$. It plays a critical role to find the
transition parameters of macroscopic quantum self-tapping.

\subsection{Numerical simulations of the MQST }

Now, we focus on the dynamic behavior which dominated by Eq.~(8)
and (9) without the time-dependent system parameters. We study the
effect parameters of the system  on the MQST with numerical method
starting form Eq.~(8) and (9).
\begin{figure}[!tbh]
\centering \rotatebox{0}{\resizebox *{18cm}{12cm}
{\includegraphics{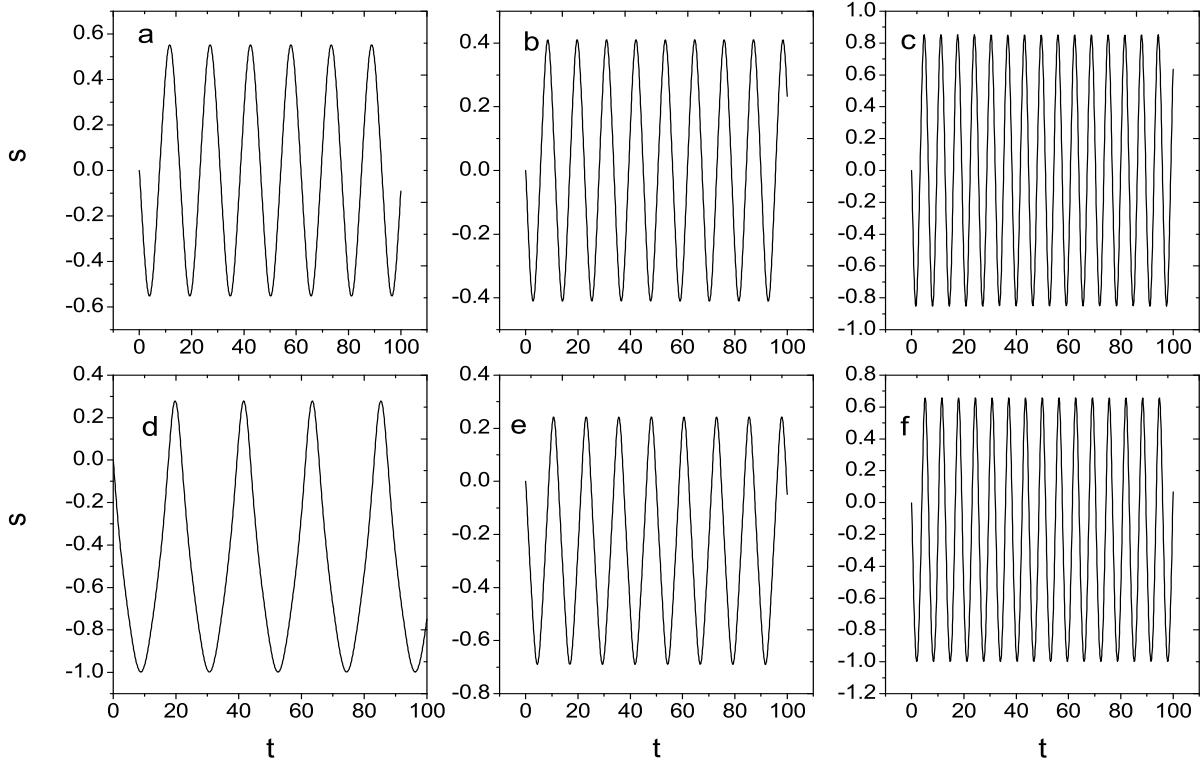}}}\caption{ The time evolution of the
relative population from Eqs.(8) and (9) with initial conditions
$s(0)=0$, $\theta_0=\pi/2$ and parameter: (a)
$c=0.1$, $\lambda=0.45$, $v=0.2$, and $\gamma=0$; (b) $c=0.1$, $\lambda=0.95$, $v=0.2$, and $\gamma=0$; (c) $c=0.1$, $\lambda=0.45$, $v=0.8$, and $\gamma=0$;
(d) $c=0.1$, $\lambda=0.45$, $v=0.2$, and $\gamma=0.5$;
(e) $c=0.1$, $\lambda=0.95$, $v=0.2$, and $\gamma=0.5$;
(f) $c=0.1$, $\lambda=0.45$, $v=0.8$, and $\gamma=0.5$;}
\end{figure}
\begin{figure}[!tbh]
\centering \rotatebox{0}{\resizebox *{14cm}{8cm}
{\includegraphics{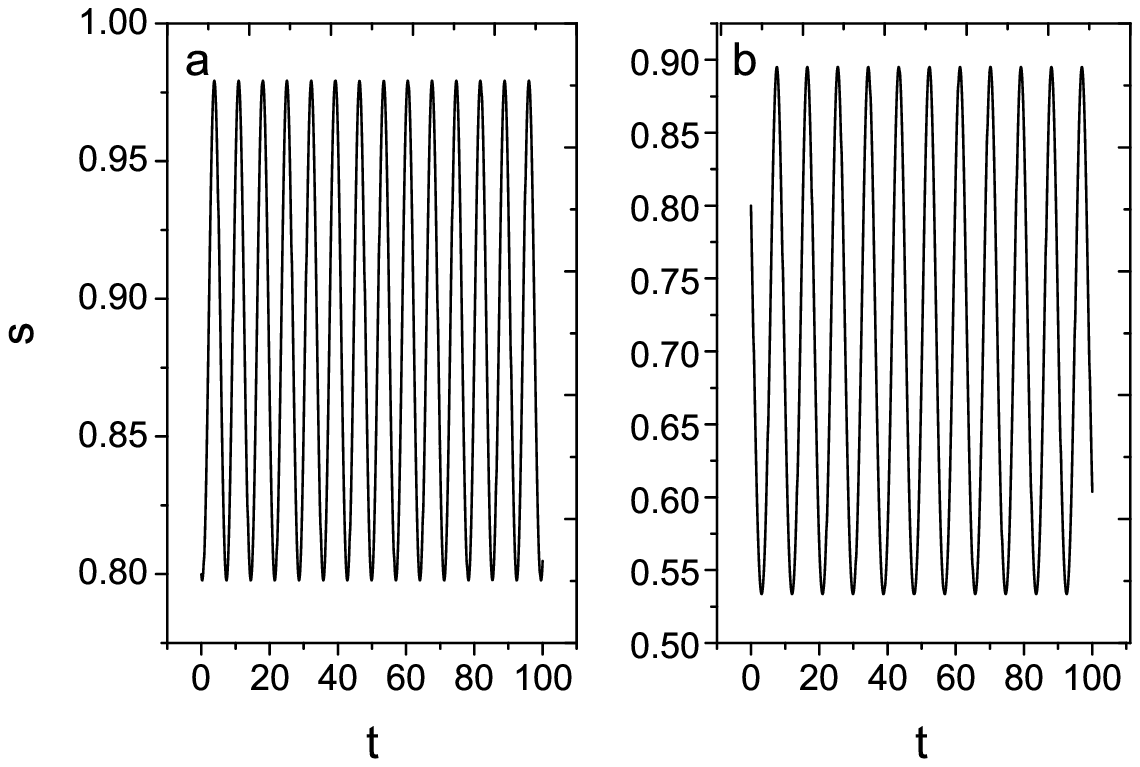}}} \caption{the time evolution of the
relative population from Eqs.~(8) and (9). (a) initial conditions
$s(0)=0.8$, $\theta_0=\pi$ (b) initial conditions $s(0)=0.8$, $\theta_0=\pi/2$, and the other parameters $c=0.1$, $\lambda=0.45$, $v=0.2$, and $\gamma=0$.}
\end{figure}

Choosing  initial condition $s(0)=0$, $\theta(0)=\pi/2$, the time
evolutions of the relative population Fig.~(3a)-(3d) show some very
absorbing features. In Fig.~3(a), the oscillations are regular and
the average the relative population $\bar{s}$ is zero for
symmetric well case ($\gamma=0$) with a special parameter, but the
corresponding MQST does not appear. If we increase $\lambda$ from
0.45 to 0.95 in Fig.~3(b), the MQST does not still appear, but the
oscillating period becomes short. Similarly, rising $\upsilon$ ,
we obtain the same result as shown in Fig.~3(c).

Here, we study impacting asymmetric well case ($\gamma\neq0$) on
the MQST. when we enhance the level bias to  $\gamma=0.5$ the
average the relative population is changed to $-0.41$ in Fig.3(d).
Correspondingly, the oscillating period of $s$ is longer and the
MQST  emerges. Note that parameter $c$, $\lambda$ and $\upsilon$
impact greatly on the MQST which are plotted in Fig.~3(e) and (f).
In fig.~3(e), when $\lambda$ is from $0.45$ to $0.95$, the MQST
is suppressed with shorter oscillating period. Similarly, with
increasing $\upsilon$, the average  relative population are
changed to $-0.21$ and the oscillating period becomes shorter
again, as seen in Fig.~3(f). Thus, the influence of parameter $c$
,$\lambda$ ,$\upsilon$ and $\gamma$ on the MQST of the system  is
very dramatic. In the case of $\gamma =0$, fixing the other
parameters and changing the initial condition from
$s(0)=0,\theta(0)=\pi/2$ of Fig.3 to
$s(0)=0.8,\theta(0)=\frac{\pi}{2}$ and $s(0)=0.8,\theta(0)=\pi$,
we observe that the MQST always emerges with varying
$s(0),\theta(0)$. The oscillating period is decreased comparing to
Fig.3(a)and Fig.~3(d), but the $\bar{s}$ is increased to
$-0.86,-0.72$ as shown in Fig.~4.

 According to the above analysis, we can draw conclusion that when
 the initial conditions $s(0)=0$, $\theta(0)=\pi/2$ are read, the
 parameter $c$, $\lambda$, $\upsilon$ can impact on the MQST for asymmetric
 well case($\gamma\neq0$). In addition, in the symmetric case, the MQST
 does not appear and those parameters only affect the oscillating
 period of the system. Besides, the initial conditions can impact
 the MQST for anyone parameter set.

 \section{Numerical simulation of chaos by applying periodic modulation on the lever bias}

As a whole, the elementary features of chaos is that the dynamic
behaviors are unpredictable for a deterministic system. It is very
sensitive for the initial conditions and parameters of the system.
So, according to these characteristics, we can adjust the
parameters to make the system get into or get out of the chaos, in
other words, we can control the regime appearing chaos. In this
section We discuss the chaotic behaviors of the system by
numerical method.

\begin{figure}[!tbh]
\centering \rotatebox{0}{\resizebox *{18cm}{12cm}
{\includegraphics{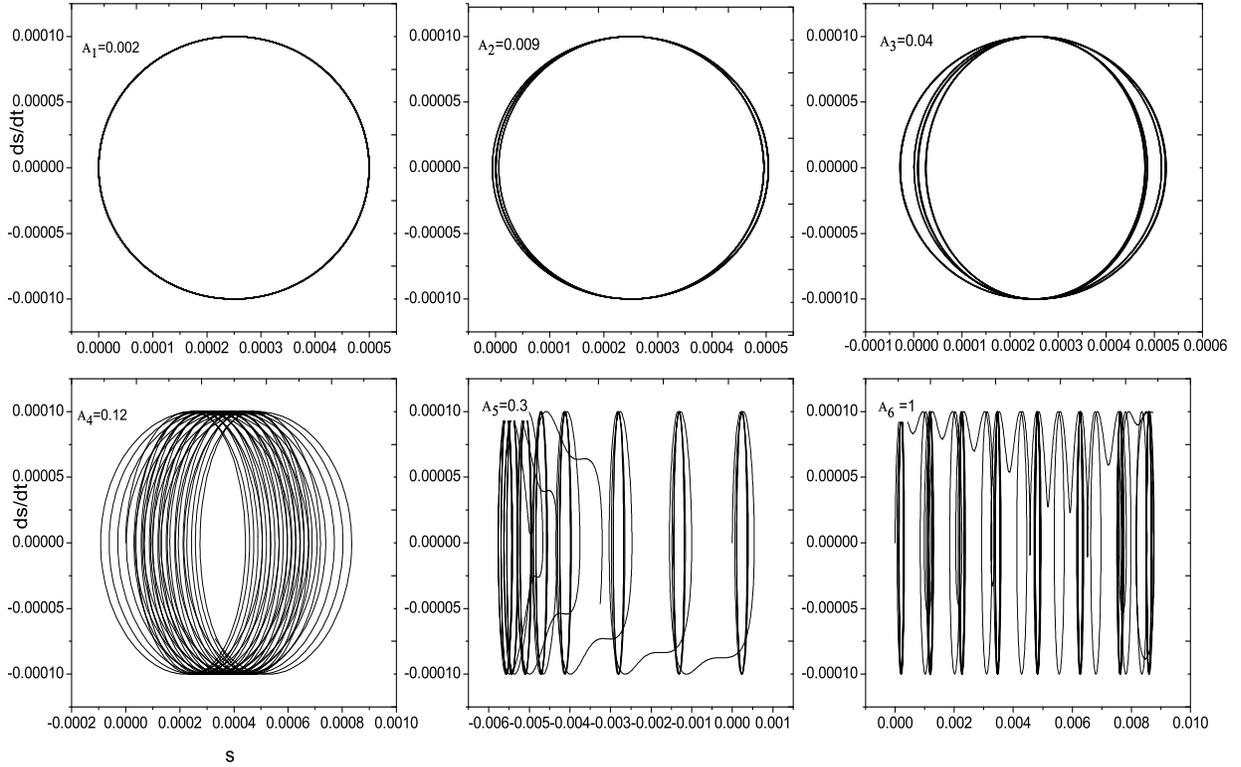}}} \caption{Dynamical phase orbits of
the dimensionless variables ($s, ds/dt$) from Eqs.~(31) and (32) with
parameters
$\upsilon=0.001$, $c=0.1$, $\lambda=0.45$, $\omega=0.1$, $s(0)=0$, $\theta_0=\pi$,
and (a) $A_1=0.002$, (b) $A_1=0.009$, (c) $A_1=0.04$, (d)
$A_1=0.12$, (e) $A_1=0.3$, (f)=$A_1=1$. Here, $A_1$ denotes the
amplitude of the time-dependent relative energy.}
\end{figure}

If we apply periodic modulation on the lever bias
$\gamma=A_0+A_1sin(\omega t)$, the chaos will appear in a special
region, where $A_0, A_1$ stand for initial phase and amplitude
respectively. Inserting this into Eqs.~(8)and (9), one derives the
below dynamic equation.
\begin{eqnarray}
\dot{s}&=&-\upsilon\sqrt{1-s^2}\sin\theta \\
\dot{\theta}&=&A_0+A_1 \sin(\omega t)+(c+2\lambda)s+\frac{\upsilon
s}{\sqrt{1-s^2}}\cos\theta
\end{eqnarray}

\begin{figure}[!tbh]
\centering \rotatebox{0}{\resizebox *{15cm}{10cm}
{\includegraphics{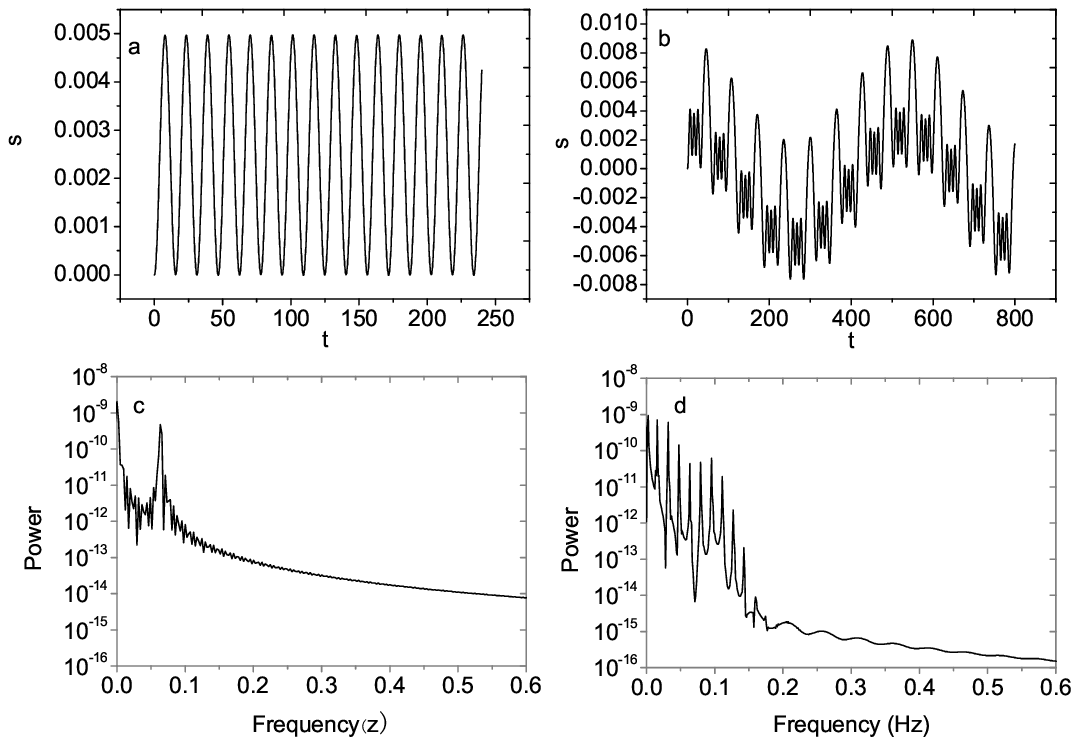}}} \caption{(a) and (b): The time
evolution of the relative population of the relative population
from Eqs.~(31)and (32) with the parameters
$\upsilon=0.001$, $A_0=0.4$, $c=0.1$, $\lambda=0.45$, $\omega=0.1$, $s(0)=0$, $\theta(0)=\pi$, and (a) $A_1=0.002$, (b) $A_1=0.3$ (c) and (d): The corresponding power
spectrum, where the parameters in Fig.~6(c)are the same with
Fig.~6(a) and the parameters in Fig.~6(d)are the same with
Fig.~6(b).}
\end{figure}

Starting from Eqs.~(32), It is found that the dynamics behavior of
the system is periodic in some special parameters region and it
will vary from order to chaos with the increase of $A_1$ , as
shown in Fig.5. With initial conditions $s(0)=0,\theta(0)=\pi$,
the phase orbit is a period-one cycle and the corresponding
oscillation is a Rabi oscillation for the set of parameters with
amplitude $A_1=0.002$, as in Fig.~5(a). In this case, we set the
oscillating period of the relative population $T$. When
$A_1=0.009$, the phase orbit becomes  period-two in Fig.~5(b). It
means the oscillating period of $s$ arriving at $2T$. Then the
phase orbit increases from that of period-four to period-eight
with increasing $A_1$ as shown in Fig.~5(c)and (d). Fig.~5(e) and
5(f) are plotted for $A_1=0.3$ and $A_1=1$, where the phase orbit
does not show a clear periodicity which signifies the emergence of
chaos.

From the above analysis, we find that the oscillating period of
the relative population varies from a period-one limit-cycle to
period-two to period-four and then to period-eight and finally all
limit-cycles tend to infinity with $\gamma$ increasing. It
exhibits a process from order to chaos, through the
period-doubling bifurcations~\cite{ce26}. That is to say, for a
set of fixed parameter $\upsilon$, $c$, $\lambda$, $A_0$, $A_1$, $s(0)$, $\theta(0)$
and $\omega$, the first-order derivative of relative population
transform from the single period to multiple period and get into
chaos at last with the increase of vibration amplitude $A_1$.

For the aim of showing the chaotic MQST, we present the plots of
the time evolution of the relative population and corresponding
plots of power spectra from Eqs.~(31) and (32) in Fig.~6. And the
parameter of Fig.5(a) is accord with Fig.~6(a) and 6(c) where the
system oscillates periodically. Making use of those parameters of
Fig.~5(e), we plot Fig.~6(b)and 6(d). It shows that the power
spectrum appears confusion and the average value of the relative
population is less than zero, which implies the existence of the
chaotic behaviors .

\section{Summary and conclusion}

In this paper, we study the stability and chaos of BEC with
repulsive two- and three-body interactions immersed in a
one-dimensional optical lattice. The stability of the steady-state
solution are analyzed with the  linear stability theorem. The
analytical results show: (1) For $\upsilon>0$ and
$c+2\lambda\geq-\upsilon$ or $\upsilon<0$ and
$c+2\lambda\leq-\upsilon$, the stability of the steady-state
solution($\theta_1=2n\pi,s_1=0$) is in the critical case. (2) For
$c+2\lambda>0$ and $\upsilon>c+2\lambda$ or $c+2\lambda<0$ and
$\upsilon>0$, the steady-state
solution($\theta_2=(2n+1)\pi,s_2=0$) is the critical stability.
(3) For $(c+2\lambda)_2>\upsilon_2$, the steady-state solution
($\theta_{3,4}=(2n+1)\pi,S_{3,4}=\pm\sqrt{1-(\frac{\upsilon}{c+2\lambda})^2}$)
is also critically stable. When these relationship are not
satisfied, the corresponding steady-state solution are unstable. A
typical tuning-fork bifurcation of steady-state relative
population appears in special parameter region. And the existence
of three-body interaction can change the bifurcation point of the
system, which is shown as Fig.~1. It plays a important role for
stability analysis of the system.

The critically stable steady-state solution indicates the
existence of the stationary MSQT. The stable behaviors of the
system change constantly with the increase of $\lambda$ and get a
general criteria for the self-trapping trajectories,
$H<-\upsilon$. In addition, we also investigate the effects of the
initial conditions, a set of parameters
{$c,\upsilon,\lambda,\gamma$} on MQST. It  shows that
$c,\upsilon,\lambda$ could affect on the MQST when
$s(0)=0,\theta_0=\pi$ for $\gamma\neq0$. Particularly, the initial
value $s(0)=0, \theta_0=\pi$  or $s(0)=0,\theta_0=\pi/2$ can
directly impact on the MQST. Finally, we discuss the chaos
behaviors by applying the modulation on the energy bias
($\gamma=A_0+A_1sin\omega t$). In this case,  the system will go
into chaos through the period-doubling bifurcations with the
increasing of $\lambda$, and the time evolution of the relative
population and power spectra indicate the existence of the chaos
MQST. It suggests that one can adjust the lasing detuning and
intensity to change the values of the parameters in experiments.
 This adjustable parameters supply the possibility for controlling
the instabilities of the system, MQST state and the chaotic
behaviors.

\begin{acknowledgments}
This work was supported by the National Natural Science Foundation of China and by the Open Project of Key Laboratory for Magnetism and Magnetic Materials of the Ministry of Education, Lanzhou University.
\end{acknowledgments}

\end{document}